\documentclass{article} 
\usepackage{iclr2027_conference,times}

\usepackage{amsmath,amsfonts,bm}

\def\eqref#1{equation~\ref{#1}}

\def\1{\bm{1}}

\DeclareMathAlphabet{\mathsfit}{\encodingdefault}{\sfdefault}{m}{sl}
\SetMathAlphabet{\mathsfit}{bold}{\encodingdefault}{\sfdefault}{bx}{n}

\usepackage{hyperref}
\usepackage{url}
\usepackage{makecell}
\usepackage{tabularx}
\usepackage{booktabs}
\usepackage{amssymb}
\usepackage{amsmath}
\usepackage{algorithm}
\usepackage{algpseudocode}
\usepackage[table]{xcolor}
\usepackage{multirow}
\usepackage{graphicx}
\usepackage{dsfont}
\usepackage{listings}
\usepackage{xcolor}
\usepackage{wrapfig}
\usepackage[most]{tcolorbox}
\tcbuselibrary{listings,breakable}
\usepackage{xcolor}

\newtcblisting{promptbox}[1]{
    enhanced,
    breakable,
    listing only,
    colback=cyan!4,
    colframe=cyan!4,
    boxrule=0pt,
    arc=2.5mm,
    left=3mm,
    right=3mm,
    top=2mm,
    bottom=2mm,
    title={#1},
    coltitle=blue!45!black,
    fonttitle=\bfseries\small\sffamily,
    colbacktitle=cyan!4,
    attach boxed title to top left={
        xshift=3mm,
        yshift=-1mm
    },
    boxed title style={
        boxrule=0pt,
        colback=cyan!4,
        colframe=cyan!4,
        left=0pt,
        right=0pt,
        top=0pt,
        bottom=0pt
    },
    listing options={
        basicstyle=\ttfamily\scriptsize,
        breaklines=true,
        breakatwhitespace=true,
        columns=fullflexible,
        keepspaces=true,
        showstringspaces=false
    }
}

\title{Asking What Matters: Disagreement-Guided Ambiguity Detection for Interactive Text-to-SQL}

\author{
\begin{tabular}{@{}l@{}}
{\bfseries
Su Dong$^{1}$, Jiayan Lin$^{2}$, Yujia Liu$^{2}$,
Zijin Hong$^{1}$, Qinggang Zhang$^{3}$, Feiran Huang$^{4}$,} \\
{\bfseries
Xiao Huang$^{1}$} \\[3pt]
{\normalfont
$^{1}$Hong Kong Polytechnic University $^{2}$Jinan University $^{3}$Jilin University $^{4}$Beihang University} \\[3pt]
{\normalfont\ttfamily
\{su.dong,zijin.hong\}@connect.polyu.hk;} \\
\end{tabular}
}

\iclrfinalcopy 
\begin{document}

\maketitle

\begin{abstract}
Interactive text-to-SQL requires an LLM agent not only to generate executable SQL, but also to decide which uncertainty to clarify within a limited interaction budget.
However, current clarification policies have three limitations: (i) they target surface ambiguity without testing its SQL impact; (ii) they miss ambiguities implicit in schema routes or domain knowledge; and (iii) they spend limited turns on unfocused or redundant questions.
These limitations expose a critical challenge: identifying SQL-impacting ambiguity before the intended SQL is known.
In this paper, we propose \textbf{\textsc{DisGuAD}}, a \textbf{\textsc{Dis}}agreement-\textbf{\textsc{Gu}}ided \textbf{\textsc{A}}mbiguity \textbf{\textsc{D}}etection framework that specializes clarification selection while leaving the downstream SQL generator unchanged.
\textsc{DisGuAD} treats disagreement among plausible SQL interpretations as direct evidence of which uncertainty matters to the final query.
Specifically, it constructs a query-term inventory, elicits contrastive SQL interpretations through complementary probe families, compares their canonical SQL components, and attributes recurring differences to user-facing clarification targets.
A lightweight prioritization and dialogue-state mechanism then selects one high-impact unresolved target for clarification.
Experiments on \textsc{BIRD-Interact} Lite and Full with two LLM backbones demonstrate consistent improvements in task success and clarification quality.
Compared with the original interaction policy, \textsc{DisGuAD} improves normalized Reward by 4.40--8.97 points and raises overall critical-ambiguity coverage from 48.62\% to 82.75\% and from 45.76\% to 85.14\%.
Its clarification signals further improve two independently designed static text-to-SQL frameworks by up to 13.97 Reward points while keeping their SQL generators fixed.
These results demonstrate that disagreement among plausible SQL queries provides actionable evidence for selecting clarification targets that matter to downstream generation.
\end{abstract}

\section{Introduction}
Generating executable SQL requires committing to decisions that users do not always state explicitly~\citep{irina2024ambrosia,ding2026ambisql,liu2026evaluating}. 
A request may leave undefined how a metric should be computed, which schema entity a phrase refers to, or what filtering, aggregation, or ranking criterion should be applied~\citep{yu2019cosql,yu2019sparc}. 
Although LLM-based text-to-SQL systems can generate increasingly complex queries~\citep{hong2025next,liu2025survey}, they are commonly formulated under the assumption that the input question fully specifies the intended semantics~\citep{yu2018spider,li2023bird}. 
This assumption breaks down when multiple reasonable interpretations lead to different SQL queries.
Interactive text-to-SQL addresses such ambiguity by allowing an LLM agent to ask clarification questions before generating the final query~\citep{huo2026birdinteract}.
This setting is no longer only about producing SQL from a given request.
The agent must also determine \emph{what to ask}: among many linguistically plausible uncertainties, it must identify the one whose resolution is most useful for the resulting SQL query~\citep{liu2025abstain}.

Existing interactive pipelines commonly rely on the LLM itself to identify ambiguity and immediately verbalize a clarification question~\citep{gao2023dailsql,huo2026birdinteract}.
Although flexible, this strategy does not explicitly test whether the selected uncertainty changes the SQL semantics.
It consequently exhibits three limitations.
(i) \textbf{Surface-level targeting:} a question may focus on an expression that appears linguistically underspecified but has little effect on projection, filtering, joins, aggregation, or other SQL decisions~\citep{irina2024ambrosia}.
(ii) \textbf{Incomplete ambiguity coverage:} consequential uncertainty is often implicit in a schema route, domain-specific metric, 

\begin{wrapfigure}{r}{0.48\textwidth}
    \vspace{-0.7\baselineskip}
    \centering
    \includegraphics[width=\linewidth]{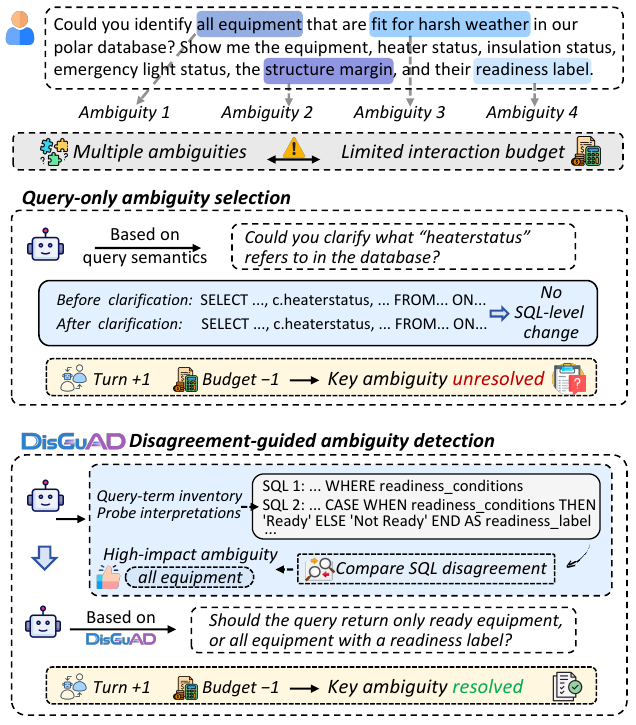}
    \vspace{-5mm}
    \caption{Ambiguity selection in interactive text-to-SQL. A query-only policy asks about \texttt{heaterstatus}, leaving the SQL unchanged and the key ambiguity unresolved. \textsc{DisGuAD} instead uses SQL disagreement to identify and resolve the high-impact result-scope ambiguity.}
    \vspace{-1mm}
    \label{fig:intro}
\end{wrapfigure}
threshold, or external definition, and may therefore be missed by clarification based only on the surface form of the request~\citep{huo2026birdinteract}.
(iii) \textbf{Unfocused interaction:} the agent may combine multiple targets in one question or revisit an ambiguity that has already been queried, consuming interaction turns without resolving additional SQL decisions~\citep{yao2019misp}.
These failures are visible in our clarification-target analysis: under two LLM backbones, the original interaction policy covers fewer than half of the benchmark-annotated critical ambiguities, with overall coverage of only 48.62\% and 45.76\%.
The limitation is therefore not simply whether the agent can ask a fluent question, but whether it can locate the uncertainty that matters to the final query.

Effective clarification poses a critical challenge: identifying \emph{SQL-impacting ambiguity} before the intended SQL is known.
The key question is not whether a phrase admits several linguistic interpretations, but whether plausible interpretations of that phrase induce different SQL decisions.
This observation suggests a direct source of evidence~\citep{wang2024macsql,li2024codes,pourreza2025chase,li2026dpc,li2026deepeye}.
If independently elicited interpretations disagree on a projection, predicate, join path, aggregation level, ranking rule, or database operation, then the assumption responsible for that disagreement is a useful clarification target.
Conversely, uncertainty that does not alter the induced SQL provides weaker evidence for interaction~\citep{farquhar2024detecting}.
We therefore treat disagreement between plausible SQL queries as an ambiguity sensor: contrastive SQL generation exposes alternative decisions, component-level comparison identifies where they diverge, and attribution maps those differences back to concepts that a user can resolve~\citep{manakul2023selfcheck,kuhn2023iclrsemantic}.

In this paper, we propose \textbf{\textsc{DisGuAD}}, a \textbf{\textsc{Dis}}agreement-\textbf{\textsc{Gu}}ided \textbf{\textsc{A}}mbiguity \textbf{\textsc{D}}etection framework for interactive text-to-SQL.
\textsc{DisGuAD} is a plug-in clarification policy that leaves the downstream SQL generator unchanged.
Given an ambiguous request, it first constructs a query-term inventory containing potential natural-language anchors.
It then elicits contrastive SQL interpretations through four complementary probe families that vary major query subgoals, relational routes, implicit definitions, and individual ambiguity axes.
\textsc{DisGuAD} canonicalizes the generated queries, identifies recurring component-level disagreements, and maps the resulting evidence back to user-facing ambiguity candidates.
A lightweight prioritization and dialogue-state mechanism selects the highest-value unresolved target and converts it into a deterministic clarification question.
Experiments under the official conversational setting of \textsc{BIRD-Interact}~\citep{huo2026birdinteract} show consistent improvements on both the Lite and Full benchmarks with two LLM backbones.
Compared with the original interaction policy, \textsc{DisGuAD} improves normalized Reward by 4.40--8.97 points and raises overall critical-ambiguity coverage from 48.62\% to 82.75\% and from 45.76\% to 85.14\% for the two backbones.
Moreover, replacing the original clarification dialogue with \textsc{DisGuAD} interaction improves two independently designed static text-to-SQL frameworks~\citep{gao2023dailsql,pourreza2023dinsql}, yielding gains of up to 13.97 Reward points while keeping their SQL generators fixed.
These results show that SQL-query disagreement provides actionable evidence for selecting clarification targets and produces interaction signals that transfer across downstream generation pipelines.

Our contributions are summarized as follows:
\begin{itemize}
    \item We identify SQL-impacting ambiguity selection as a central challenge in interactive text-to-SQL and formulate disagreement between plausible SQL queries as direct evidence of which uncertainty should be clarified.
    \item We propose \textsc{DisGuAD}, a plug-in framework that combines query-term extraction, contrastive SQL interpretation probes, component-level disagreement analysis, and user-facing ambiguity attribution without modifying the final SQL generator.
    \item Extensive experiments on \textsc{BIRD-Interact} demonstrate consistent improvements across two benchmark variants and LLM backbones. Further analysis shows substantially higher critical-ambiguity coverage and confirms that \textsc{DisGuAD} interaction generalization.
\end{itemize}

\section{Related Work}
Traditional text-to-SQL methods formulate the task as single-turn semantic parsing~\citep{wang2020rat,li2023resdsql}. 
Recent LLM-based systems further employ task decomposition, in-context example selection, candidate generation, and self-correction~\citep{pourreza2023dinsql,qu2024before,xie2025open,li2025omnisql,lei2025spider2,hong2026errorllm}. 
Interactive semantic parsing introduced model-based agents that decide when and where to request user feedback~\citep{yao2019misp}. 
Recent work studies ambiguity more directly: AmbiQT~\citep{bhaskar2023benchmarking} and AMBROSIA~\citep{irina2024ambrosia} associate ambiguous utterances with multiple plausible SQL interpretations, AmbiSQL~\citep{ding2026ambisql} detects ambiguity using a fine-grained taxonomy and elicits multiple-choice clarification, and BIRD-Interact evaluates multi-turn agents that interact with users and database environments~\citep{huo2026birdinteract}. 
In parallel, consistency-based LLM methods compare sampled generations to estimate semantic uncertainty or detect unreliable content~\citep{kuhn2023iclrsemantic,manakul2023selfcheck}.
\textsc{DisGuAD} instead uses component-level SQL disagreement to select user-facing clarification targets; extended comparisons appear in the Appendix~\ref{app:related_work}.

\section{Methods}
\label{sec:method}

\subsection{Preliminaries and Overview}
\label{sec:preliminaries}
\label{sec:method_overview}

\paragraph{Task Formulation.}
Given a natural language user question $q$ and a database $\mathcal{D}$ with schema $\mathcal{S}$, the conventional text-to-SQL task learns a mapping $f$ converting the question into SQL query:
\begin{equation}
\label{eq:text_to_sql}
f:(q,\,\mathcal{S})\rightarrow y,
\end{equation}
where $y$ is the SQL query that expresses the intent of $q$. 
This formulation assumes that the user question fully specifies the intended query. 
Interactive text-to-SQL relaxes this assumption by presenting the agent $\mathcal{M}_{\alpha}$ with an ambiguous question $\tilde{q}$. 
The agent interacts with a user $\mathcal{U}$ to acquire the missing information before generating the final SQL query.
We define the environment as $\mathcal{E}=(\mathcal{D},\mathcal{S},\mathcal{K})$, where $\mathcal{K}$ denotes the available external knowledge, and let $H_t$ denote the dialogue history through turn $t$.
At each non-final turn, the agent asks a question $q^{(\textsc{Ask})}_{t}$, receives a response (observation) $o_t$, and updates the history for subsequent interaction:
\begin{equation}
\label{eq:interactive_text_to_sql}
q^{(\textsc{Ask})}_{t}=\mathcal{M}^{(\textsc{Ask})}_{\alpha}\bigl(\tilde{q},\,\mathcal{E},\,H_{t-1}\bigr),\,o_t = \mathcal{U}\bigl(q^{(\textsc{Ask})}_{t},\,H_{t-1}\bigr),\,H_t = H_{t-1}\oplus(q^{(\textsc{Ask})}_{t},\,o_t).
\end{equation}
where $\oplus$ appends the new question--response pair and $H_0=\varnothing$. 
At the terminal turn $T$, the agent produces the final PostgreSQL query $\widehat{y}=\mathcal{M}^{(\textsc{SQL})}_{\alpha}(\tilde{q},\,\mathcal{E},\,H_{T-1})$ at turn $T$. The superscripts distinguish the question-generation and SQL-generation modes of the same agent with parameters $\alpha$.

\paragraph{\textsc{DisGuAD} Overview.}
Our proposed framework \textsc{DisGuAD} is a plug-in clarification policy for the interactive setting above. 
\textsc{DisGuAD} specializes clarification selection while leaving the underlying agent $\mathcal{M}_{\alpha}$ and its final SQL generator unchanged. 
Its central idea is to represent ambiguity through \emph{SQL query disagreement}: if two plausible interpretations of the ambiguous question $\tilde{q}$ induce different SQL decisions, the uncertainty responsible for that difference is a useful clarification target. 
\textsc{DisGuAD} exposes such differences with contrastive SQL probes and maps them back to concepts that the user can clarify. 
The complete workflow can be formalized as:
\begin{equation}
\label{eq:workflow}
q_t^{(\textsc{Ask})} = \operatorname{\textsc{DisGuAD}}\bigl(\tilde{q},\,\mathcal{S},\,\mathcal{K},\,H_{t-1}\bigr).
\end{equation}
\textsc{DisGuAD} first elicits contrastive SQL interpretations of $\tilde{q}$, identifies the semantic decisions on which these interpretations disagree, and maps the supported disagreements back to concepts that can be clarified by the user. 
Finally, it selects the highest-priority unresolved ambiguity and verbalizes it using a deterministic question template. The following sections instantiate each component.

\subsection{Query-term Inventory}
\label{sec:query_term_inventory}

The query-term inventory defines the natural-language anchors available for clarification. We use an LLM-based extractor to identify short spans in $\tilde{q}$ and assign each span a coarse ambiguity type:
\begin{equation}
\label{eq:term_inventory}
\mathcal{X}=\textsc{Ext}(\tilde{q},\,\mathcal{S},\,\mathcal{K};\widetilde{\mathcal{X}})
=\{(x_k,c_k)\}_{k=1}^{M},\, c_k \in \mathcal{C}^{(\textsc{NL})},
\end{equation}
where $x_k$ is a short text span in $\tilde{q}$ that may serve as a
clarification anchor, and $c_k$ denotes its coarse ambiguity type.
The auxiliary set $\widetilde{\mathcal{X}}$ contains heuristically
identified candidates supplied as weak extraction evidence, while
$\mathcal{C}^{(\textsc{NL})}$ comprises lexical, syntactic, semantic,
schema-linking, query-intent, and knowledge-linking ambiguity types.
These heuristic candidates improve extraction coverage but do not
determine which term should ultimately be queried.
Thus, $\mathcal{X}$ identifies potential clarification anchors, whose
SQL relevance is determined by the subsequent disagreement analysis.

\begin{figure*}[!t]
    \centering
    \includegraphics[width=\linewidth]{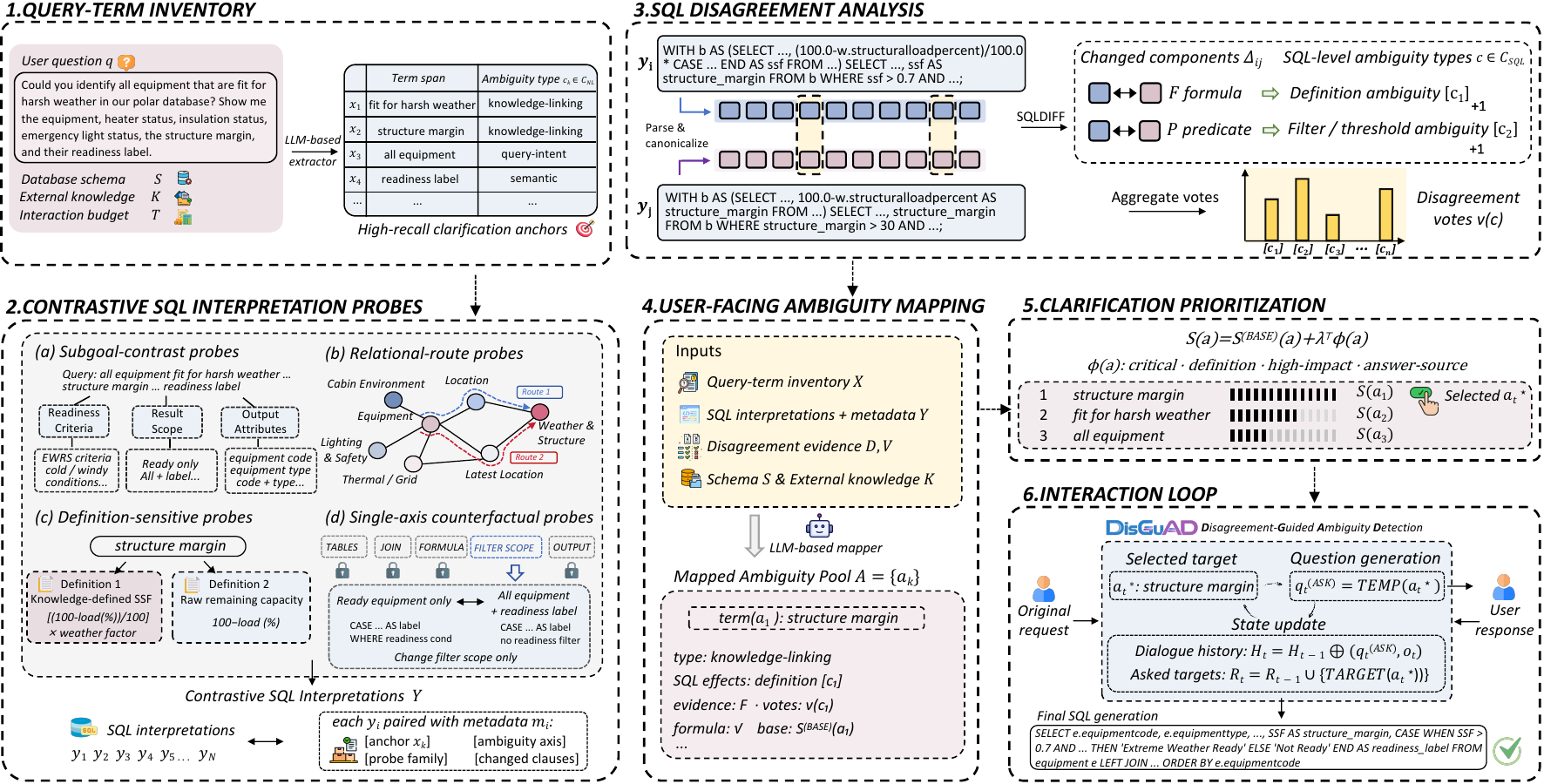}
    \vspace{-5mm}
    \caption{Overview of \textsc{DisGuAD}. It extracts clarification anchors, elicits contrastive SQL interpretations, and aggregates disagreement. The evidence is mapped to user-facing terms, prioritized against the dialogue state, and converted into a clarification question before SQL generation.}
    \vspace{-4mm}
    \label{fig:main}
\end{figure*}

\subsection{Contrastive SQL Interpretation Probes}
\label{sec:contrastive_sql_probes}

To expose alternative SQL decisions, \textsc{DisGuAD} elicits $N$ plausible PostgreSQL interpretations through four controlled probe families $\mathcal{P}=\{\texttt{SUB},\,\texttt{REL},\,\texttt{DEF},\,\texttt{CF}\}$. 
This controlled elicitation is designed to reduce incidental variation caused by repeatedly sampling from one single prompt. 
For the $i$-th probe, let $p_i\in\mathcal{P}$ denote its probe family. We generate a plausible SQL query together with metadata describing the assumptions under which it was produced:
\begin{equation}
\label{eq:contrastive_probe}
(y^{(\textsc{Prb})}_i,m_i)=\textsc{Prb}_{p_i}(\tilde{q},\,\mathcal{S},\,\mathcal{K},\,
\mathcal{X}),\,
\end{equation}
where $y^{(\textsc{Prb})}_i$ is a plausible PostgreSQL interpretation and $m_i$ is the metadata describing its underlying assumptions, including the probe family, ambiguity axis, anchored span, expected changed clauses, referenced schema elements, and relevant knowledge identifiers.

The four probe families expose complementary sources of query-level variation: \textbf{subgoal contrast} (\texttt{SUB}) varies major query decisions, including projections, tables, predicates, aggregations, ordering, limits, and database operations; \textbf{relational route} (\texttt{REL}) explores alternative table paths, joins, grouping levels, filter scopes, and final projections; \textbf{definition sensitivity} (\texttt{DEF}) tests alternative definitions of scores, rates, ratios, and other domain-specific metrics; and \textbf{single-axis counterfactuals} (\texttt{CF}) vary one ambiguity axis while holding the remaining SQL decisions fixed.

Across these families, $m_i$ traces each query variation back to the assumption that produced it. The probe-generated SQL queries serve only as diagnostic evidence for the subsequent disagreement analysis; none is selected or reranked as the final answer.

\subsection{SQL Disagreement Analysis}
\label{sec:sql_disagreement}

\paragraph{Component-level Disagreement.}
Each probe-generated SQL query is parsed into the decisions that define its execution semantics. 
These decisions include the database operation (e.g., \texttt{SELECT}, \texttt{UPDATE}, or \texttt{DELETE}), projected attributes, referenced tables and join paths, filtering predicates, derived formulas, grouping keys, \texttt{HAVING} conditions, ordering criteria, and row limits. 
We canonicalize each component before comparison so that formatting, aliases, and other surface-level variations do not create spurious disagreements. For each pair of probe outputs, we then define:
\begin{equation}
\label{eq:pairwise_disagreement}
\Delta_{ij} =  \operatorname{\textsc{SQLDiff}}\left(y_i^{(\textsc{Prb})},\,y_j^{(\textsc{Prb})}\right),\,1\leq i<j\leq N.
\end{equation}
The operator returns the canonical SQL components on which the two interpretations disagree. 
Thus, $\Delta_{ij}$ captures structural disagreement between the SQL queries rather than textual variation.

\paragraph{Disagreement Voting.}
We group component-level differences into a predefined set of SQL
ambiguity types $\mathcal{C}^{(\textsc{SQL})}$, including projection,
schema or join, filter or threshold, definition or granularity, ranking,
and database-operation ambiguity. The operator
$\operatorname{\textsc{Type}}(\Delta_{ij})$ returns the subset of these ambiguity
types indicated by the changed components in $\Delta_{ij}$. Because a
difference observed in only one probe pair may be incidental, we define
the disagreement vote for each type
$c\in\mathcal{C}^{(\textsc{SQL})}$ as
\begin{equation}
\label{eq:disagreement_vote}
v(c) = \sum_{1\leq i<j\leq N} \mathds{1}\left[c\in\operatorname{\textsc{Type}}(\Delta_{ij})\right].
\end{equation}
Here, $\mathds{1}[\cdot]$ is the indicator function, and $v(c)$ counts the number of probe pairs that disagree on ambiguity type $c$. A larger $v(c)$ provides stronger disagreement evidence that the corresponding ambiguity affects the resulting SQL query;
smaller value suggests that the ambiguity may be incidental.

\subsection{Mapping SQL Disagreement to User-Facing Ambiguity}
\label{sec:disagreement_mapping}

Component-level disagreements identify unstable SQL decisions, but clause names and schema elements cannot be presented directly as clarification questions. \textsc{DisGuAD} therefore uses an LLM-based mapper to attribute the disagreement evidence to the clarification anchors in $\mathcal{X}$:
\begin{equation}
\label{eq:disagreement_mapping}
\mathcal{A} = \textsc{Map}\Bigl(\mathcal{X},\,\bigl\{(y_i^{(\textsc{Prb})}, m_i)\bigr\}_{i=1}^{N},\,\bigl\{\Delta_{ij}\bigr\}_{i<j},\,\bigl\{v(c)\bigr\}_{c\in\mathcal{C}^{(\textsc{SQL})}},\,\mathcal{S},\mathcal{K}\Bigr)=\{a_\ell\}_{\ell=1}^{L}.
\end{equation}
Here, $\mathcal{A}$ is the resulting pool of user-facing ambiguity
candidates. Each $a_\ell\in\mathcal{A}$ represents a clarification target grounded in the original request and records its ambiguity type, affected SQL decisions, supporting probe interpretations, disagreement vote, definition requirements, and base priority.

The mapper attributes observed changes in projections, predicates, joins, grouping, ordering, limits, and database operations to the interpretation of an anchor in $\mathcal{X}$. 
This attribution is particularly important for knowledge-intensive expressions such as ``stability score'', ``release decision rule'', or ``risk indicator'', whose SQL realizations may depend on a formula, threshold, JSON path, or external definition rather than a single schema attribute.
Terms grounded in $\mathcal{K}$ may also be considered when $\tilde{q}$ refers to the corresponding concept. 
SQL artifacts such as clause names, table aliases, and generated column aliases are excluded, while generic spans such as ``ID'', ``data'' or ``result'' are retained only when the disagreement evidence reveals a concrete schema or projection conflict.

\subsection{Clarification Prioritization and Dialogue State}
\label{sec:clarification_selection}

\paragraph{Ambiguity Prioritization.}
\textsc{DisGuAD} assigns each mapped ambiguity $a\in\mathcal{A}$ a priority score:
\begin{equation}
\label{eq:disguad_priority}
\operatorname{S}(a) = \operatorname{S}^{(\textsc{Base})}(a) + \boldsymbol{\lambda}^{\top}\boldsymbol{\phi}(a),
\end{equation}
where $\operatorname{S}(a)$ is the target score and $\operatorname{S}^{(\textsc{Base})}(a)$ is the mapper-assigned base priority; $\boldsymbol{\phi}(a)\in\{0,1\}^{4}$ encodes whether $a$ is critical, definition-dependent, high-impact, or associated with an identifiable answer source, and $\boldsymbol{\lambda}\in\mathbb{R}_{\geq 0}^{4}$ contains the weights to control the contribution of these properties.

\paragraph{Dialogue State.}
To avoid repeated questions, $\mathcal{R}_t$ records the normalized
clarification targets queried through turn $t$, with
$\mathcal{R}_0=\varnothing$.
Let $\operatorname{\textsc{Target}}(a)$ denote the normalized clarification target associated with $a$.
The eligible ambiguity set at turn $t$ is:
\begin{equation}
\label{eq:eligible_candidates}
\mathcal{A}_t = \left\{ a\in\mathcal{A} \;\middle|\;\operatorname{\textsc{Askable}}(a) \ \land\ \operatorname{\textsc{Target}}(a)\notin\mathcal{R}_{t-1} \right\},
\end{equation}
where $\operatorname{\textsc{Askable}}(a)$ indicates whether $a$ can be verbalized as a user-answerable question, and $\operatorname{\textsc{Target}}$ returns its normalized clarification target.
When $\mathcal{A}_t$ is nonempty, \textsc{DisGuAD} selects the highest-priority candidate and converts it into a clarification question:
\begin{equation}
\label{eq:ambiguity_selection}
a_t^\star = \operatorname*{arg\,max}_{a\in\mathcal{A}_t} \operatorname{S}(a),\, q_t^{(\textsc{Ask})} = \operatorname{\textsc{Temp}}(a_t^\star).
\end{equation}
The template $\operatorname{\textsc{Temp}}(\cdot)$ requests a definition or formula for definition ambiguities, an ordering criterion or row limit for ranking ambiguities, an intended entity or attribute for schema ambiguities, and the intended action for operation ambiguities.
If no mapped ambiguity is eligible, \textsc{DisGuAD} first selects an unqueried anchor from $\mathcal{X}$, which is ranked below all disagreement-supported candidates. If neither source provides an eligible target, question generation falls back to the underlying agent $\mathcal{M}_{\alpha}^{(\textsc{Ask})}$. After a mapped or inventory-based question is issued, its normalized target is added to $\mathcal{R}_t$. Under the benchmark protocol, the resulting question is emitted as \texttt{<s>[QUESTION]</s>}.

\section{Experiments}
\subsection{Experimental Setups}
\label{sec:experimental_setup}

\paragraph{Dataset.}
We use \textsc{BIRD-Interact}~\citep{huo2026birdinteract}, a dynamic interactive text-to-SQL benchmark that evaluates SQL generation through multi-turn interaction. Each task consists of an ambiguous priority question followed by a context-dependent follow-up question in a PostgreSQL environment with database metadata and hierarchical external knowledge. We evaluate our method on both \textsc{BIRD-Interact-Full} and \textsc{BIRD-Interact-Lite} under the official $c$-\textsc{Interact} setting, where the system follows a predefined clarification protocol before submitting the final SQL query.

\paragraph{Evaluation Metrics.}
Following \textsc{BIRD-Interact}, we report \textbf{Priority Question Success Rate (PQ-SR)}, \textbf{Follow-up Question Success Rate (FQ-SR)}, and \textbf{Normalized Reward (Reward)}. PQ-SR measures the success rate on the initial priority question, while FQ-SR measures the success rate on the context-dependent follow-up question. A SQL query is counted as successful only if it passes all executable test cases. Reward is the official normalized task-level score on a 0--100 scale, jointly measuring performance across the priority and follow-up questions.

\paragraph{Compared Methods.}
We compare two groups of methods. 
For interactive text-to-SQL, \textbf{Vanilla} directly generates SQL without clarification, \textbf{\textsc{BIRD-Interact}} uses the benchmark's standard interaction pipeline, and \textbf{\textsc{DisGuAD}} replaces the default ambiguity-selection policy with disagreement-guided clarification. 
To evaluate transferability, we inject clarification signals into two representative static text-to-SQL frameworks, \textbf{DAIL-SQL}~\citep{gao2023dailsql} and \textbf{DIN-SQL}~\citep{pourreza2023dinsql}. 
For each framework, we compare \textbf{Vanilla}, \textbf{w/ Original}, and \textbf{w/ \textsc{DisGuAD}}, where the latter two add the clarification dialogue from the original policy or \textsc{DisGuAD}. 
Because the downstream SQL generator is fixed, this comparison isolates the value of the clarification signal.

\paragraph{Implementation Details.}
In all main experiments, \textsc{DisGuAD} generates $K=8$ SQL candidates for each ambiguous user query. The candidate budget is evenly distributed across four complementary reasoning paths. We instantiate \textsc{DisGuAD} with two LLM backbones: DeepSeek-V4-Flash and GPT-5.6-Luna. For each experiment, we use the same backbone for the system model, user simulator, and ambiguity module to avoid confounding effects from cross-model mismatch. All LLM calls are made through the official model APIs. Experiments are conducted in the conversational setting using the official \textsc{BIRD-Interact} Docker evaluator for evaluation, together with the PostgreSQL database containers provided by the benchmark. Local orchestration, database management, and evaluation scripts are run on a machine equipped with six NVIDIA RTX 3090 GPUs.

\begin{table*}[t]
\centering
\small
\setlength{\tabcolsep}{3.5pt}
\caption{
Main results on \textsc{BIRD-Interact-Lite} and \textsc{BIRD-Interact-Full}. 
We report Priority Question Success Rate (PQ-SR), Follow-up Question Success Rate (FQ-SR), and normalized Reward. 
The upper block evaluates \textsc{DisGuAD} as an interactive text-to-SQL method, while the lower block evaluates whether \textsc{DisGuAD}-selected signals transfer to static text-to-SQL frameworks.
Best results within each comparable group are in bold, and second-best results are underlined.
Parenthesized gains are measured against the corresponding Original setting.
}
\vspace{2mm}
\label{tab:main_lite_full_results}

\resizebox{\textwidth}{!}{
\begin{tabular}{llcccccc}
\toprule
\multirow{2}{*}{\textbf{Framework}} &
\multirow{2}{*}{\textbf{Setting}}
& \multicolumn{3}{c}{\textbf{\textsc{Lite}}} 
& \multicolumn{3}{c}{\textbf{\textsc{Full}}} \\
\cmidrule(lr){3-5} \cmidrule(lr){6-8}
& 
& \textbf{PQ-SR} $\uparrow$ & \textbf{FQ-SR} $\uparrow$ & \textbf{Reward} $\uparrow$
& \textbf{PQ-SR} $\uparrow$ & \textbf{FQ-SR} $\uparrow$ & \textbf{Reward} $\uparrow$ \\
\midrule

\rowcolor{gray!15}
\multicolumn{8}{c}{\textit{DeepSeek-V4-Flash}} \\
\midrule

Vanilla & No interaction 
& 16.00 & 9.00 & 13.90
& 15.83 & 9.50 & 13.93 \\
\textsc{BIRD-Interact} & Original
& \underline{43.00} & \underline{28.00} & \underline{38.50}
& \underline{29.00} & \underline{16.50} & \underline{25.25} \\
\textsc{DisGuAD} & \textsc{DisGuAD}
& \textbf{50.67}{\scriptsize(+7.67)} 
& \textbf{31.33}{\scriptsize(+3.33)} 
& \textbf{44.87}{\scriptsize(+6.37)}
& \textbf{33.83}{\scriptsize(+4.83)} 
& \textbf{21.00}{\scriptsize(+4.50)} 
& \textbf{29.98}{\scriptsize(+4.73)} \\

\midrule
\midrule

DAIL-SQL & Vanilla
& 10.00 & 5.33 & 8.60
& 10.67 & 6.17 & 9.32 \\
DAIL-SQL & w/ Original
& \underline{48.67} & \underline{31.33} & \underline{43.47}
& \underline{42.67} & \underline{26.83} & \underline{37.92} \\
DAIL-SQL & w/ \textsc{DisGuAD}
& \textbf{52.33}{\scriptsize(+3.66)}
& \textbf{32.33}{\scriptsize(+1.00)}
& \textbf{46.33}{\scriptsize(+2.86)}
& \textbf{44.50}{\scriptsize(+1.83)}
& \textbf{27.83}{\scriptsize(+1.00)}
& \textbf{39.50}{\scriptsize(+1.58)} \\

\midrule

DIN-SQL & Vanilla
& 10.00 & 5.00 & 8.50
& 11.83 & 7.17 & 10.43 \\
DIN-SQL & w/ Original
& \underline{50.67} & \underline{32.00} & \underline{45.07}
& \underline{41.83} & \underline{25.67} & \underline{36.98} \\
DIN-SQL & w/ \textsc{DisGuAD}
& \textbf{55.00}{\scriptsize(+4.33)}
& \textbf{34.00}{\scriptsize(+2.00)}
& \textbf{48.70}{\scriptsize(+3.63)}
& \textbf{45.00}{\scriptsize(+3.17)}
& \textbf{28.50}{\scriptsize(+2.83)}
& \textbf{40.05}{\scriptsize(+3.07)} \\

\midrule
\rowcolor{gray!15}
\multicolumn{8}{c}{\textit{GPT-5.6-Luna}} \\
\midrule

Vanilla & No interaction
& 17.00 & 9.00 & 14.60
& 18.17 & 10.17 & 15.77 \\
\textsc{BIRD-Interact} & Original
& \underline{30.00} & \underline{18.33} & \underline{26.50}
& \underline{21.67} & \underline{12.17} & \underline{18.82} \\
\textsc{DisGuAD} & \textsc{DisGuAD}
& \textbf{40.67}{\scriptsize(+10.67)}
& \textbf{23.33}{\scriptsize(+5.00)}
& \textbf{35.47}{\scriptsize(+8.97)}
& \textbf{26.67}{\scriptsize(+5.00)}
& \textbf{15.17}{\scriptsize(+3.00)}
& \textbf{23.22}{\scriptsize(+4.40)} \\

\midrule
\midrule

DAIL-SQL & Vanilla
& 10.33 & 4.00 & 8.43
& 13.17 & 8.33 & 11.72 \\
DAIL-SQL & w/ Original
& \underline{31.67} & \underline{18.67} & \underline{27.77}
& \underline{21.50} & \underline{14.67} & \underline{19.45} \\
DAIL-SQL & w/ \textsc{DisGuAD}
& \textbf{47.33}{\scriptsize(+15.66)}
& \textbf{27.67}{\scriptsize(+9.00)}
& \textbf{41.43}{\scriptsize(+13.66)}
& \textbf{33.17}{\scriptsize(+11.67)}
& \textbf{19.67}{\scriptsize(+5.00)}
& \textbf{29.12}{\scriptsize(+9.67)} \\

\midrule

DIN-SQL & Vanilla
& 10.33 & 5.33 & 8.83
& 14.33 & 8.50 & 12.58 \\
DIN-SQL & w/ Original
& \underline{31.00} & \underline{17.67} & \underline{27.00}
& \underline{20.67} & \underline{13.17} & \underline{18.42} \\
DIN-SQL & w/ \textsc{DisGuAD}
& \textbf{46.67}{\scriptsize(+15.67)}
& \textbf{27.67}{\scriptsize(+10.00)}
& \textbf{40.97}{\scriptsize(+13.97)}
& \textbf{33.33}{\scriptsize(+12.66)}
& \textbf{20.67}{\scriptsize(+7.50)}
& \textbf{29.53}{\scriptsize(+11.11)} \\

\bottomrule
\end{tabular}
}
\vspace{-5mm}
\end{table*}

\subsection{Main Results}
\label{sec:main_results}

Table~\ref{tab:main_lite_full_results} reports the main results on \textsc{BIRD-Interact-Lite} and \textsc{BIRD-Interact-Full}. 
The upper block compares \textsc{DisGuAD} with Vanilla and the standard \textsc{BIRD-Interact} policy. 
Across both datasets and both LLM backbones, \textsc{DisGuAD} achieves the best PQ-SR, FQ-SR, and Reward. 
Compared with Original, \textsc{DisGuAD} improves Reward by 6.37/4.73 points with DeepSeek-V4-Flash and by 8.97/4.40 points with GPT-5.6-Luna on \textsc{Lite}/\textsc{Full}. 
It also largely outperforms Vanilla, whose Reward remains much lower under both backbones. 
These results show that ambiguous text-to-SQL cannot be reliably solved from the original query alone, and that selecting SQL-impactful clarification targets is crucial for final SQL quality.

The lower block evaluates whether \textsc{DisGuAD}'s clarification signals can benefit static text-to-SQL frameworks. 
We use DAIL-SQL and DIN-SQL as representative static frameworks and compare Vanilla, w/ Original, and w/ \textsc{DisGuAD}. 
Adding clarification context improves over Vanilla in all settings, and replacing Original with \textsc{DisGuAD} further improves Reward in every case. 
The gains range from +1.58 to +3.63 points with DeepSeek-V4-Flash and from +9.67 to +13.97 points with GPT-5.6-Luna. 
Since the downstream SQL generator is fixed within each comparison, these gains isolate the benefit of \textsc{DisGuAD}'s ambiguity-selection strategy rather than a stronger SQL decoder.

Overall, \textsc{DisGuAD} improves the interactive policy itself and provides reusable clarification signals for external SQL generators, supporting its role as a plug-in ambiguity-selection layer for interaction-augmented text-to-SQL.

\subsection{Clarification Target Analysis}
\label{sec:clarification_target_analysis}

We analyze whether \textsc{DisGuAD} identifies better clarification targets using the golden ambiguity annotations of benchmark. 
Clarification questions are automatically matched to golden ambiguity terms by normalized string matching and token overlap.
We report five metrics: Cov@3 and Cov@All measure coverage of SQL-impacting critical ambiguity terms within the first three questions and all questions, respectively; Know. Cov@3 measures early coverage of knowledge-related ambiguity; Single-target measures the fraction of questions that match exactly one golden ambiguity term; and Repeat measures redundant clarification.

\begin{table}[t]
\centering
\small
\setlength{\tabcolsep}{3.5pt}
\caption{
Clarification-target quality of \textsc{DisGuAD} and the original policy
against gold annotations on \textsc{BIRD-Interact-Full}. Best results per backbone are bold.
}
\label{tab:clarification_target_quality}
\vspace{2mm}
\resizebox{\columnwidth}{!}{
\begin{tabular}{llccccc}
\toprule
\textbf{Backbone} & \textbf{Method}
& \textbf{Cov@3}
& \textbf{Cov@All}
& \textbf{Know. Cov@3}
& \textbf{Single-target}
& \textbf{Repeat $\downarrow$} \\
\midrule
DeepSeek-V4-Flash & Original
& 48.43 & 48.62 & 47.44 & 38.31 & 7.38 \\
DeepSeek-V4-Flash & \textsc{DisGuAD}
& \textbf{66.06} & \textbf{82.75} & \textbf{54.68} & \textbf{44.05} & \textbf{6.14} \\
\midrule
GPT-5.6-Luna & Original
& 45.38 & 45.76 & 49.61 & 37.94 & 11.17 \\
GPT-5.6-Luna & \textsc{DisGuAD}
& \textbf{67.60} & \textbf{85.14} & \textbf{54.13} & \textbf{45.60} & \textbf{5.85} \\
\bottomrule
\end{tabular}
}
\vspace{-5mm}
\end{table}

Table~\ref{tab:clarification_target_quality} shows that \textsc{DisGuAD} improves all clarification-target metrics across both backbones. 
Most notably, Cov@All increases from 48.62\% to 82.75\% with DeepSeek-V4-Flash and from 45.76\% to 85.14\% with GPT-5.6-Luna, showing that \textsc{DisGuAD} covers substantially more SQL-impacting ambiguity targets. 
The gains in Know. Cov@3 and Single-target further indicate that \textsc{DisGuAD} better identifies knowledge-related ambiguities and produces more focused clarification questions, while the lower Repeat rate shows less redundant clarification.

\begin{figure}[t]
    \centering
    \includegraphics[width=0.95\textwidth]{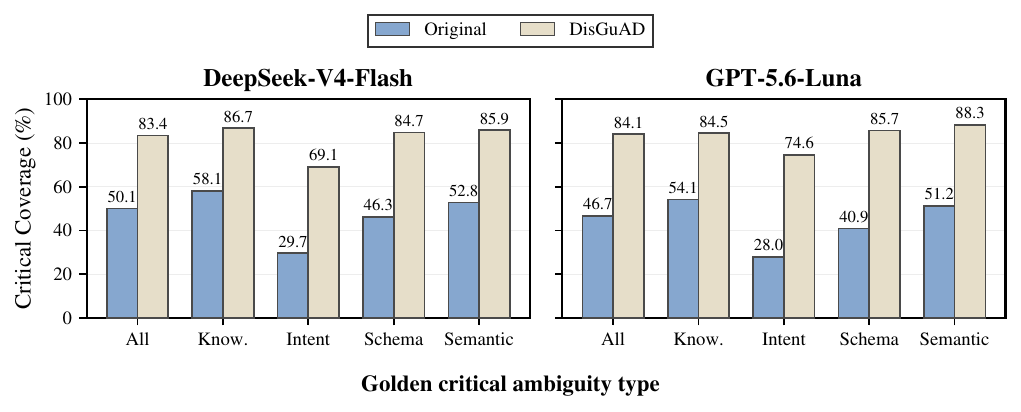}
    \vspace{-5mm}
    \caption{
    Golden ambiguity coverage on \textsc{BIRD-Interact-Full}. 
    \textsc{DisGuAD} improves SQL-impacting target coverage across both backbones; ``Know.'' denotes knowledge-linking ambiguity.
    }
    \label{fig:coverage_by_type}
\end{figure}

Figure~\ref{fig:coverage_by_type} further shows that the improvement is consistent across major ambiguity types, including knowledge-linking, intent, schema-linking, and semantic ambiguity. 
This supports the central motivation of \textsc{DisGuAD}: SQL disagreement among plausible interpretations helps identify user-facing ambiguity targets that are more likely to affect concrete SQL decisions.

\subsection{Sensitivity to Candidate Budget}
\label{sec:candidate_budget_sensitivity}

\begin{figure}[t]
    \centering
    \includegraphics[width=\columnwidth]{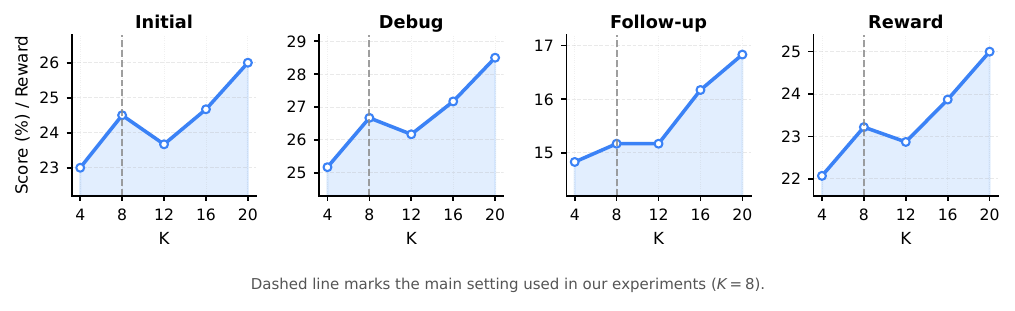}
    \vspace{-8mm}
    \caption{
    SQL-candidate budget sensitivity on \textsc{BIRD-Interact-Full}. 
    Larger $K$ generally improves all metrics; the dashed line marks the main setting $K=8$.
    }
    \label{fig:candidate_budget_sensitivity}
\end{figure}

Figure~\ref{fig:candidate_budget_sensitivity} studies the effect of the SQL-candidate budget $K$ on \textsc{BIRD-Interact-Full}. 
Increasing $K$ from 4 to 20 improves Initial success from 23.00\% to 26.00\%, Debug success from 25.17\% to 28.50\%, Follow-up success from 14.83\% to 16.83\%, and Reward from 22.07 to 25.00, indicating that a larger candidate budget exposes more diverse SQL-level disagreement signals and helps \textsc{DisGuAD} select more useful clarification targets. 
The gains are not strictly monotonic, as $K=12$ slightly underperforms $K=8$ on several metrics, suggesting that candidate quality and diversity also matter. 
This supports the use of SQL candidates as ambiguity sensors rather than final-answer candidates. 
We use $K=8$ in the main experiments as a practical trade-off, since larger budgets can further improve performance but require additional LLM calls for candidate generation.

\subsection{Ablation Study}
\label{sec:ablation}

We conduct ablation studies to understand the contribution of the main components in \textsc{DisGuAD}. 
All ablations are conducted on \textsc{BIRD-Interact-Full} with GPT-5.6-Luna as the backbone. 
We consider two groups of ablations. 
The first group removes LLM-assisted ambiguity modules, including the query-term inventory extractor and the disagreement-to-ambiguity mapper. 
The second group removes one contrastive SQL interpretation probe family at a time. 
Following the main results, we report PQ-SR, FQ-SR, and normalized Reward.

\begin{wraptable}[12]{r}{0.60\textwidth}
\vspace{-8pt}
\centering
\footnotesize
\setlength{\tabcolsep}{2pt}
\renewcommand{\arraystretch}{0.92}

\caption{Ablation study with GPT-5.6-Luna.}
\label{tab:ablation}
\vspace{2mm}
\begin{tabular}{@{}l@{\hspace{5pt}}ccc@{}}
\toprule
\textbf{Variant} & \textbf{PQ-SR} & \textbf{FQ-SR} & \textbf{Reward} \\
\midrule
Full \textsc{DisGuAD}
& \textbf{26.67} & \textbf{15.17} & \textbf{23.22} \\
\midrule
w/o Query-term Inventory
& 23.33 & 13.33 & 20.33 \\
w/o Disagreement Mapping
& 24.50 & 14.50 & 21.50 \\
\midrule
w/o Subgoal-contrast
& 26.17 & 14.83 & 22.77 \\
w/o Relational-route
& 25.83 & 15.00 & 22.58 \\
w/o Definition-sensitive
& 25.67 & 14.67 & 22.37 \\
w/o Single-axis CF
& 26.00 & 15.00 & 22.70 \\
\bottomrule
\end{tabular}
\end{wraptable}

Table~\ref{tab:ablation} shows that both LLM-assisted ambiguity modules are important. 
Removing the query-term inventory decreases PQ-SR from 26.67\% to 23.33\% and Reward from 23.22 to 20.33. 
This indicates that extracting user-facing ambiguity anchors is necessary for connecting clarification to terms that users can understand and answer. 
Removing the disagreement mapping module also causes a clear drop, reducing PQ-SR to 24.50\% and Reward to 21.50. 
This confirms that SQL-level disagreement alone is insufficient; \textsc{DisGuAD} needs the mapping step to attribute program differences back to askable ambiguity targets.

The probe-family ablations show that the contrastive SQL interpretation probes provide complementary evidence. 
Removing subgoal-contrast probes or relational-route probes lowers Reward from 23.22 to 22.77 and 22.58, respectively. 
This suggests that decomposing the query into SQL subgoals and exploring alternative relational routes both help expose ambiguity that affects the final SQL program. 
The ablations on definition-sensitive and single-axis counterfactual probes show a similar pattern, reducing Reward to 22.37 and 22.70, respectively. 
These results indicate that formula-oriented interpretations and controlled counterfactual variations provide useful additional disagreement signals. 
Overall, the full \textsc{DisGuAD} model achieves the best Reward, supporting the effectiveness of combining user-facing term extraction, SQL-disagreement attribution, and diverse contrastive SQL probes.

\section{Conclusion}
In this paper, we propose \textsc{DisGuAD}, a plug-in clarification policy that identifies SQL-impacting ambiguity through disagreement among plausible SQL interpretations. 
\textsc{DisGuAD} elicits contrastive queries, compares their canonical components, and attributes recurring differences to user-facing clarification targets while leaving the downstream SQL generator unchanged. 
Experiments on \textsc{BIRD-Interact} demonstrate consistent improvements in PQ-SR, FQ-SR, and Reward across two LLM backbones.
Clarifications selected by \textsc{DisGuAD} also improve two widely used text-to-SQL frameworks, demonstrating that its interaction signals transfer across downstream SQL generators.
Ablation studies further confirm the contributions of query-term extraction, disagreement mapping, and controlled SQL probes. 
These results establish SQL disagreement as actionable evidence for deciding what an interactive text-to-SQL agent should ask.

\section*{AI Use Statement}

In this work, we used generative AI tools to assist with manuscript writing and code development. Specifically, generative AI was used for language editing, improving readability, and drafting or revising parts of the manuscript, as well as for assisting with code writing and debugging. We did not use generative AI tools for research methodology design, hypothesis formulation, or experimental design. All AI-assisted text and code were reviewed and verified by the authors. We take full responsibility for the final content of this work, including all text, code, claims, and results.

\bibliography{reference}
\bibliographystyle{iclr2027_conference}

\appendix
\section{Extended Related Work}
\label{app:related_work}

\paragraph{Text-to-SQL Generation.}
Text-to-SQL systems have progressed from specialized semantic parsers to LLM-based generation through improved schema linking, structured decoding, task decomposition, and self-correction~\citep{zhong2017seq2sql,wang2020rat,
scholak2021picard,li2023resdsql,gao2023dailsql}.
Recent methods also generate and compare multiple SQL candidates~\citep{pourreza2025chase,li2026dpc}. Unlike candidate-selection approaches, \textsc{DisGuAD} uses alternative queries to identify what should be clarified rather than to choose the final SQL.

\paragraph{Conversational and Interactive Text-to-SQL.}
Conversational benchmarks evaluate context-dependent SQL generation over pre-collected dialogue sequences~\citep{yu2019sparc,yu2019cosql}, whereas interactive parsers actively request structured or free-form user feedback~\citep{yao2019misp,li2020mean,elgohary2020speak}.
BIRD-Interact further evaluates dynamic interaction with users and database environments~\citep{huo2026birdinteract}. \textsc{DisGuAD} complements these settings by selecting one unresolved, SQL-impacting ambiguity at each interaction turn.

\paragraph{Ambiguity in Text-to-SQL.}
AmbiQT~\citep{bhaskar2023benchmarking} and AMBROSIA~\citep{irina2024ambrosia} associate ambiguous requests with multiple plausible interpretations and SQL queries, while AmbiSQL resolves detected ambiguity through multiple-choice clarification~\citep{ding2026ambisql}.
Rather than relying only on surface ambiguity or a predefined taxonomy,\textsc{DisGuAD} identifies clarification targets from recurring differences among plausible SQL queries.

\paragraph{Disagreement and Uncertainty in LLM Generation.}
Consistency-based methods compare sampled LLM generations to estimate semantic uncertainty, detect unsupported content, or diagnose uncertainty sources~\citep{kuhn2023iclrsemantic,manakul2023selfcheck,
chen2024quantifying,nan2025multiple,liu2025abstain}.
These methods operate over free-form responses.
\textsc{DisGuAD} instead compares structured SQL components and attributes their disagreement to user-facing clarification targets.

\section{Case Study}
\label{sec:case_study}

We present a representative case from the \texttt{polar\_2} database to illustrate how \textsc{DisGuAD} identifies SQL-impacting ambiguity targets. 
The user asks:

\begin{quote}
\small
\textit{
Could you identify all equipment that are fit for harsh weather in our polar database? 
Show me the equipment, heater status, insulation status, emergency light status, the structure margin, and their readiness label.
}
\end{quote}

This request contains several SQL-impacting ambiguities, including the definition of ``fit for harsh weather'', the formula for ``structure margin'', the intended output identifier for ``equipment'', and whether ``readiness label'' means filtering only ready equipment or assigning a label to all relevant equipment.

\begin{table}[t]
\centering
\small
\setlength{\tabcolsep}{3pt}
\caption{
Case study on the \texttt{polar\_2} database. 
The original interaction asks clarification questions but fails to target the SQL-impacting ambiguities, while \textsc{DisGuAD} asks about ambiguity targets that change filtering, joins, computed expressions, and output labeling.
}
\label{tab:case_study_polar}
\resizebox{\columnwidth}{!}{
\begin{tabular}{p{0.25\columnwidth}p{0.32\columnwidth}p{0.36\columnwidth}}
\toprule
\textbf{Ambiguity} & \textbf{Original behavior} & \textbf{\textsc{DisGuAD} behavior} \\
\midrule
\textit{fit for harsh weather} 
& Does not clarify the full readiness rule and applies implicit filters. 
& Clarifies the Extreme Weather Ready rule: SSF $>0.7$, heater not Off, insulation not Poor, and emergency light On or Testing. \\
\midrule
\textit{structure margin} 
& Uses a threshold without clarifying the formula. 
& Clarifies it as Structural Safety Factor rounded to two decimals. \\
\midrule
\textit{all equipment} 
& Interprets the request as returning only equipment satisfying readiness conditions. 
& Includes each structurally evaluable equipment record and assigns a readiness label. \\
\midrule
\textit{readiness label} 
& Treats readiness as a filtering criterion. 
& Treats readiness as a classification output: \texttt{Extreme Weather Ready} or \texttt{Not Ready}. \\
\midrule
Join semantics
& Uses inner joins, excluding equipment with missing cabin, lighting, or insulation records. 
& Uses left joins, so missing auxiliary records do not remove equipment from the output population. \\
\bottomrule
\end{tabular}
}
\end{table}

The original interaction policy asks clarification questions, but they do not resolve the ambiguity targets that determine the final SQL semantics. 
As shown in Table~\ref{tab:case_study_polar}, the resulting SQL still makes two incorrect implicit assumptions. 
First, it treats readiness as a filtering condition, returning only equipment that satisfies the harsh-weather criteria. 
Second, it uses inner joins to cabin, thermal, and lighting tables, which removes equipment records when auxiliary status information is missing. 
As a result, the query loses the intended semantics of ``all equipment'' and ``readiness label'': the expected behavior is to retain each structurally evaluable equipment record and classify it as either \texttt{Extreme Weather Ready} or \texttt{Not Ready}.

\textsc{DisGuAD} instead decomposes the uncertainty into user-facing clarification targets. 
It asks about the definition of ``fit for harsh weather'', the formula for ``structure margin'', the meaning of ``readiness label'', whether ``all equipment'' should include non-ready equipment, and which equipment identifier should be returned. 
After receiving these clarifications, \textsc{DisGuAD} generates SQL that uses left joins to preserve the output population and a \texttt{CASE} expression to assign the readiness label. 
This separates inclusion criteria from classification criteria, producing the correct final SQL.

This case highlights that \textsc{DisGuAD} improves clarification by identifying ambiguity terms whose resolution changes concrete SQL decisions, including filtering, join type, computed expressions, and output labeling. 
The comparison also shows why asking a clarification question is not sufficient by itself: the question must target the ambiguity that actually affects the SQL program.

\section{Prompt Templates for \textsc{DisGuAD}}
\label{app:prompt_templates}

This appendix reports the prompt templates used by \textsc{DisGuAD} to decide clarification questions. 
The templates correspond to the main components in Section~\ref{sec:method}: query-term inventory construction, contrastive SQL interpretation probes, mapping SQL disagreement to user-facing ambiguity, and template-based clarification question realization. 
\textsc{DisGuAD} does not modify the downstream final SQL generation prompt in the main interactive experiments. 
Instead, it changes the clarification policy by selecting which ambiguity should be asked before final SQL synthesis.

We use braces to denote runtime placeholders, such as \texttt{\{user\_query\}}, \texttt{\{database\_schema\}}, and \texttt{\{external\_knowledge\}}. 
The prompts are formatted for readability while preserving the logic used in our implementation.

\subsection{Query-term Inventory Prompt}
\label{app:query_term_inventory_prompt}

This prompt instantiates the query-term inventory construction described in Section~\ref{sec:query_term_inventory}. 
The goal is to extract user-facing spans that may later serve as clarification anchors. 
The extractor does not decide the final question by itself; it only provides a natural-language search space for subsequent SQL-disagreement attribution.

\begin{promptbox}{System Prompt}
You extract user-facing ambiguity terms for interactive text-to-SQL.
Return only JSON. Do not use hidden benchmark labels.
Keep rationales concise.
\end{promptbox}

\begin{promptbox}{User Prompt Template}
Database:
{database_name}

Original ambiguous user request:
{user_query}

Database schema:
{database_schema}

Visible external knowledge:
{external_knowledge}

Heuristic term candidates for reference only:
{heuristic_terms}

Extract user-facing terms or phrases that may need clarification
before generating SQL.

Use these coarse ambiguity categories:
- Lexical Ambiguity: a token has multiple meanings in context.
- Syntactic Ambiguity: a phrase can attach to different parts
  of the request.
- Semantic Ambiguity: wording is vague or underspecified.
- Schema Linking Ambiguity: a query term could map to multiple
  schema elements.
- Query Intent Ambiguity: operation, ranking, filtering,
  aggregation, limit, or DDL/DML intent is unclear.
- Knowledge Linking Ambiguity: a concept may link to an external
  definition or formula, but the link is implicit or unclear.

Important:
- Prefer spans that appear verbatim in the user query.
- Include metric or formula terms such as score, index, rate,
  ratio, percentage, coefficient, or factor.
- Include scope terms such as latest, recent, top, first, or batch.
- Include action terms such as create, update, or manage when
  DDL/DML semantics may matter.
- Include schema-like terms such as ID, status, type, name,
  facility, or telescope when mapping may be ambiguous.
- Do not include generic stopword spans.
- Do not reveal or use benchmark golden ambiguity labels.
- Do not copy the heuristic list mechanically; use it only as
  weak evidence.

Return valid JSON with the following structure:
{
  "terms": [
    {
      "term_id": "t1",
      "span": "...",
      "normalized_term": "...",
      "coarse_category_candidates": ["Semantic Ambiguity"],
      "why_potentially_ambiguous": "...",
      "possible_sql_effects": [
        "select expression",
        "where predicate"
      ],
      "priority_hint": "critical | useful | low"
    }
  ]
}
\end{promptbox}

\subsection{Contrastive SQL Interpretation Probes}
\label{app:contrastive_sql_probe_prompts}

This prompt instantiates the contrastive probing stage described in Section~\ref{sec:contrastive_sql_probes}. 
Given the ambiguous request, schema, external knowledge, and query-term inventory, \textsc{DisGuAD} elicits multiple plausible SQL interpretations. 
These interpretations are not used for reranking or final SQL selection. 
Instead, they serve as diagnostic probes for exposing SQL-level disagreement. 
In the main experiments, we set the candidate budget to $K=8$.

\begin{promptbox}{Probe Families}
1. Subgoal-contrast probes (SUB)
   Decompose the user request into major SQL subgoals, including
   output columns, tables and joins, filters, metric or formula,
   aggregation grain, ordering or limit, and database operation.
   Generate plausible SQL interpretations that differ because
   one subgoal has a different interpretation.

2. Relational-route probes (REL)
   Explore alternative relational realizations of the same request,
   including table paths, join paths, grouping levels, filter scopes,
   and final projections when multiple schema routes are plausible.

3. Definition-sensitive probes (DEF)
   Focus on metric, score, index, rate, ratio, percentage,
   coefficient, and factor phrases. Generate interpretations that
   explore different formulas, knowledge links, thresholds, or
   numerator/denominator choices.

4. Single-axis counterfactual probes (CF)
   Change one ambiguity axis while keeping other SQL decisions as
   stable as possible. Possible axes include metric formula,
   schema entity, join path, aggregation grain, filter scope,
   ranking or limit, output projection, database operation,
   and null handling.
\end{promptbox}

\begin{promptbox}{System Prompt}
You generate plausible PostgreSQL interpretations for ambiguous
text-to-SQL tasks. Return only JSON. Do not execute SQL.
Generate interpretations that are intentionally diverse along
user-facing ambiguity axes.
\end{promptbox}

\begin{promptbox}{User Prompt Template}
Database:
{database_name}

Original ambiguous user request:
{user_query}

Database schema:
{database_schema}

Visible external knowledge:
{external_knowledge}

Query-term inventory:
{query_terms}

Clarification memory:
{clarification_memory}

Generate exactly {num_candidates} plausible PostgreSQL interpretations
using the following contrastive probe plan:
{probe_plan}

Important rules:
- Every SQL interpretation must be plausible for the original
  user request.
- Each probe family should explore a different ambiguity source.
- Single-axis counterfactual probes should primarily change exactly
  one ambiguity axis while keeping other SQL decisions stable.
- Prefer schema-valid SQL using visible tables and columns.
- Do not invent hidden benchmark labels or unavailable external
  knowledge.
- Anchor the main assumption of each interpretation to a user-facing
  query span or visible knowledge term.
- Keep SQL concise but complete enough for clause-level comparison.

Return valid JSON with the following structure:
{
  "candidates": [
    {
      "candidate_id": "sub_1",
      "probe_family": "SUB",
      "primary_axis": "metric_formula",
      "anchored_query_span": "...",
      "coarse_category": "Semantic Ambiguity",
      "sql_effect_type_hint": "projection_or_metric_ambiguity",
      "assumption": "...",
      "sql": "SELECT ...",
      "expected_changed_clauses": ["select"],
      "used_tables": ["..."],
      "used_columns": ["..."],
      "used_knowledge_ids": [],
      "unresolved_terms": ["..."],
      "confidence": 0.7
    }
  ]
}
\end{promptbox}

\subsection{Mapping SQL Disagreement to User-Facing Ambiguity}
\label{app:disagreement_mapping_prompt}

This prompt corresponds to the disagreement mapping stage in Section~\ref{sec:disagreement_mapping}. 
The input consists of the query-term inventory, probe-generated SQL interpretations, candidate metadata, SQL-difference evidence, and ambiguity-type vote counts. 
The goal is to attribute SQL-level disagreement to user-facing terms that can be asked about in natural language.

This step treats SQL differences as evidence, not as direct clarification targets. 
For example, \textsc{DisGuAD} should not ask the user whether a \texttt{GROUP BY} clause should change. 
Instead, it should ask about the natural-language term that explains why plausible SQL interpretations differ in grouping, filtering, projection, join path, ordering, or formula computation.

\begin{promptbox}{System Prompt}
You map ambiguity evidence to user-facing query terms for
interactive text-to-SQL. Return only JSON.
Use query terms as anchors. Do not reveal hidden benchmark labels.
\end{promptbox}

\begin{promptbox}{User Prompt Template}
Original ambiguous user request:
{user_query}

Query-term inventory:
{query_terms}

User-query ambiguity categories already detected:
{query_term_categories}

Visible external knowledge:
{external_knowledge}

Clarification memory and interaction history:
{clarification_memory}

SQL interpretation metadata and SQL:
{sql_candidates}

SQL difference evidence:
{sql_diff}

SQL-level ambiguity summary derived from candidate differences:
{sql_ambiguity_summary}

SQL-difference ambiguity type votes:
{prioritized_sql_ambiguities}

Query-first heuristic mappings:
{heuristic_mappings}

Task:
Reason from observed SQL disagreement back to likely user-facing
causes. For each important SQL-level ambiguity, decide which
original-query term, previous-answer term, or visible knowledge term
most likely caused that SQL disagreement.

Use the query-term inventory as the primary anchor. SQL difference
evidence is evidence for ambiguity, not a source of askable terms.

Coarse categories:
- Lexical Ambiguity
- Syntactic Ambiguity
- Semantic Ambiguity
- Schema Linking Ambiguity
- Query Intent Ambiguity
- Knowledge Linking Ambiguity

Rules:
- Prefer terms or spans from the original query.
- You may include a knowledge term from the user's previous answers
  if it appears in memory and needs follow-up.
- You may include a visible external-knowledge metric name only when
  the original query appears to refer to that concept indirectly.
- Do not ask about terms already asked in memory unless the user
  answer introduced a new knowledge term.
- Do not use whole sentences as terms.
- Never use SQL syntax or implementation artifacts as terms:
  SELECT, FROM, WHERE, JOIN, CASE, WHEN, THEN, ELSE, TRUE, FALSE,
  NULL, NULLIF, COALESCE, CAST, JSON_AGG, CTE, alias names, table
  aliases, parentheses, operators, or literal SQL keywords.
- Downrank low-value terms such as "ID", "score", "result", or
  "data" unless candidate evidence shows a real schema or projection
  conflict and the term appears in the user query.
- If SQL differences are real but cannot be anchored to a query term,
  answer term, or visible knowledge term, mark them as
  unanchored_sql_noise and do not make them askable.
- Prefer terms that explain high-vote SQL ambiguity types and would
  materially change SELECT, WHERE, GROUP BY, JOIN, ORDER BY, LIMIT,
  or DDL/DML behavior.
- A good mapping should explain why this user term caused this
  observed SQL ambiguity, not merely repeat a candidate assumption.
- The best ambiguity should be the one most likely to change
  the final SQL.

Return valid JSON with the following structure:
{
  "ambiguities": [
    {
      "term": "...",
      "source": "original_query | user_answer | visible_knowledge | unanchored_sql_noise",
      "coarse_category": "Knowledge Linking Ambiguity",
      "ambiguity_type": "knowledge_linking_ambiguity",
      "sql_effect_type": "projection_or_metric_ambiguity",
      "caused_sql_ambiguity_types": [
        "aggregation_or_metric_ambiguity"
      ],
      "matched_sql_votes": 5,
      "query_span": "...",
      "question_need": "...",
      "candidate_ids": ["c1", "c2"],
      "changed_sql_parts": ["select", "columns"],
      "candidate_interpretations": ["...", "..."],
      "evidence_examples": ["..."],
      "reason": "...",
      "importance": "critical | useful | low",
      "askable": true,
      "requires_user": true,
      "requires_formula": true,
      "confidence": 0.75,
      "priority_score": 0.8
    }
  ]
}
\end{promptbox}

\subsection{Template-Based Clarification Question Realization}
\label{app:question_realization_templates}

After selecting an ambiguity target, \textsc{DisGuAD} realizes the final clarification question using deterministic templates. 
We use templates rather than an additional LLM question rewriter in the main experiments. 
This separates the decision of \emph{which ambiguity to ask about} from the surface form of the question.

Let \texttt{term} denote the selected ambiguity target and \texttt{sql\_effect} denote its predicted SQL effect type.

\begin{promptbox}{Main Templates}
If the ambiguity requires a formula:
  Could you clarify the exact definition or formula for {term}?

If the SQL effect is related to sorting or ranking:
  Could you clarify the intended ordering, ranking, or limit for {term}?

If the SQL effect is related to schema linking:
  Could you clarify which entity or attribute you mean by {term}?

If the SQL effect is related to DDL/DML behavior:
  Could you clarify the intended create, insert, update, or upsert
  behavior for {term}?

If a specific question_need is available:
  Could you clarify {term}? {question_need}

Otherwise:
  Could you clarify what you mean by {term}?
\end{promptbox}

\begin{promptbox}{Fallback Slot Questions}
metric_formula:
  Are there any remaining metric, score, or formula definitions
  that should change the SQL?

filter_threshold:
  Are there any remaining filters, thresholds, or inclusion rules
  that should change the SQL?

output_shape:
  Are there any remaining output columns, aliases, or JSON keys
  that should change the SQL?

grouping_grain:
  Should the results use any grouping grain beyond what you
  already specified?

ranking_limit:
  Are there any remaining ordering, ranking, or row-limit
  requirements for the final result?

default:
  Are all SQL-relevant requirements now complete, or is there
  one remaining detail I should use?
\end{promptbox}

\begin{promptbox}{Interaction Format}
<s>{clarification_question}</s>
\end{promptbox}

\section{Prompt Scope and Reproducibility Notes}
\label{app:prompt_scope}

We report only the prompt templates used by \textsc{DisGuAD} to decide clarification questions. 
The appendix excludes prompts that are not part of \textsc{DisGuAD}'s ambiguity-selection policy, including the original \textsc{BIRD-Interact} system prompt, the user simulator prompt, downstream final SQL generation prompts, DAIL-SQL and DIN-SQL prompts, answer-observation prompts, canonical-plan prompts, SQL validation prompts, and SQL repair prompts.

All prompt templates are fixed across the reported experiments. 
Within each experiment, the same LLM backbone is used for the system model, user simulator, and \textsc{DisGuAD} ambiguity module. 
Across backbone comparisons, only the LLM backbone changes.

\end{document}